\documentclass[prb,aps,twocolumn,showpacs,amsmath,amssymb]{revtex4-1}
\usepackage{graphicx}
\usepackage{dcolumn}
\usepackage{color}
\usepackage{bm}
\usepackage{units}
\usepackage{natbib}

\begin{document}

\author{V.~Fritsch$^{1}$,
 N.~Bagrets$^1$ , G.~Goll$^1$, W. Kittler$^1$, M.~J.~Wolf$^1$, K.~Grube$^2$, C.-L. Huang$^{1,2}$ and H.~v.~L\"{o}hneysen$^{1,2}$}

\affiliation{$^{1}$Karlsruher Institut f\"{u}r Technologie, Physikalisches Institut, 76131~Karlsruhe, Germany \\
$^{2}$Karlsruher Institut f\"{u}r Technologie, Institut f\"{u}r Festk\"{o}rperphysik, 76344 Karlsruhe, Germany}

\date{\today}

\title{Approaching quantum criticality in a partially geometrically frustrated heavy-fermion metal}

\begin{abstract}
In the antiferromagnetic (AF) heavy-fermion system  \mbox{CePdAl} the magnetic Ce ions form a network of equilateral triangles in the $(001)$ plane, similar to the kagom\'{e} lattice, with one third of the Ce moments not participating in the long-range order. The N\'{e}el temperature $T_{\textrm{N}} = 2.7\,\unit{K}$ is reduced upon replacing Pd by Ni in CePd$_{1-x}$Ni$_{x}$Al, with $T_{\textrm{N}} \rightarrow 0$ for $x = 0.144$, where the specific heat $C$ exhibits a $C/T \propto - \log\,T$ dependence. Within the Hertz-Millis-Moriya model of quantum criticality, this behavior might indicate $2$D critical antiferromagnetic fluctuations arising from the decoupling of  $3$D magnetic order by frustration. On the other hand, the simultaneous presence of Kondo effect and  geometric frustration might entail a new route to quantum criticality.
\end{abstract}

\maketitle

\section{Introduction}
Quantum phase transitions occurring strictly speaking only at absolute zero temperature, are found  in insulating and metallic systems, and also in ultracold dilute atomic gases in optical lattices.\cite{Sachdev2011,Gegenwart2008,Greiner2002} The common feature  of these very different material classes lies in the fact that the competition between low-energy scales can be tuned by a nonthermal parameter, such as pressure, magnetic or electric field, and chemical composition for the condensed-matter systems. In heavy-fermion materials, the strong exchange $J$ between $f$-electrons and conduction electrons can lead to quenching of the $f$-electron-derived (nearly) localized magnetic moments via the Kondo effect or, if $J$ becomes weaker, to long-range magnetic order via the Ruderman-Kittel-Kasuya-Yosida (RKKY) interaction mediated by the conduction electrons.\cite{Doniach1977}

Quantum critical behavior has been observed in a number of heavy-fermion (HF) systems.\cite{Lohneysen2007} In the canonical HF system CeCu$_{6-x}$Au$_x$ where a magnetic-nonmagnetic quantum phase transition (QPT) can be tuned by composition or pressure, quasi-two-dimensional ($2$D) magnetic fluctuations arise out of $3$D long-range antiferromagnetic (AF) order, as shown by inelastic neutron scattering.\cite{Stockert1998} Furthermore, the QPT in this system is associated with unusual energy/temperature scaling of the dynamical susceptibility arising from critical fluctuations,\cite{Schroder1998,Schroder2000} also observed in UCu$_{5-x}$Al$_{x}$.\cite{Aronson1995} These findings prompted the development of alternative models of quantum criticality \cite{Coleman2001,Si2001} going beyond  the Landau-Ginzburg-Wilson model of classical second-order phase transitions and its extension to QPTs at zero temperature by Hertz,\cite{Hertz1976} Millis,\cite{Millis1993} and Moriya\cite{Moriya1995} (HMM model). These models were further nurtured by a comprehensive set of experiments on YbRh$_2$Si$_2$ that showed a drastic change of the Hall effect at a field driven QPT, incompatible with the HMM model.\cite{Paschen2004}
In the model of local quantum criticality \cite{Si2001} $2$D fluctuations are a prerequisite for the validity of the model. The issue of dimensional crossover in metallic quantum-critical magnets was discussed theoretically by Garst {\em et al.}.\cite{Garst2008} Recently, several quantum-critical systems of reduced dimensionality were compared with cubic Ce$_3$Pd$_{20}$Si$_6$.\cite{Custers2012}

\begin{figure}
\begin{center}
\includegraphics[clip,width=\columnwidth]{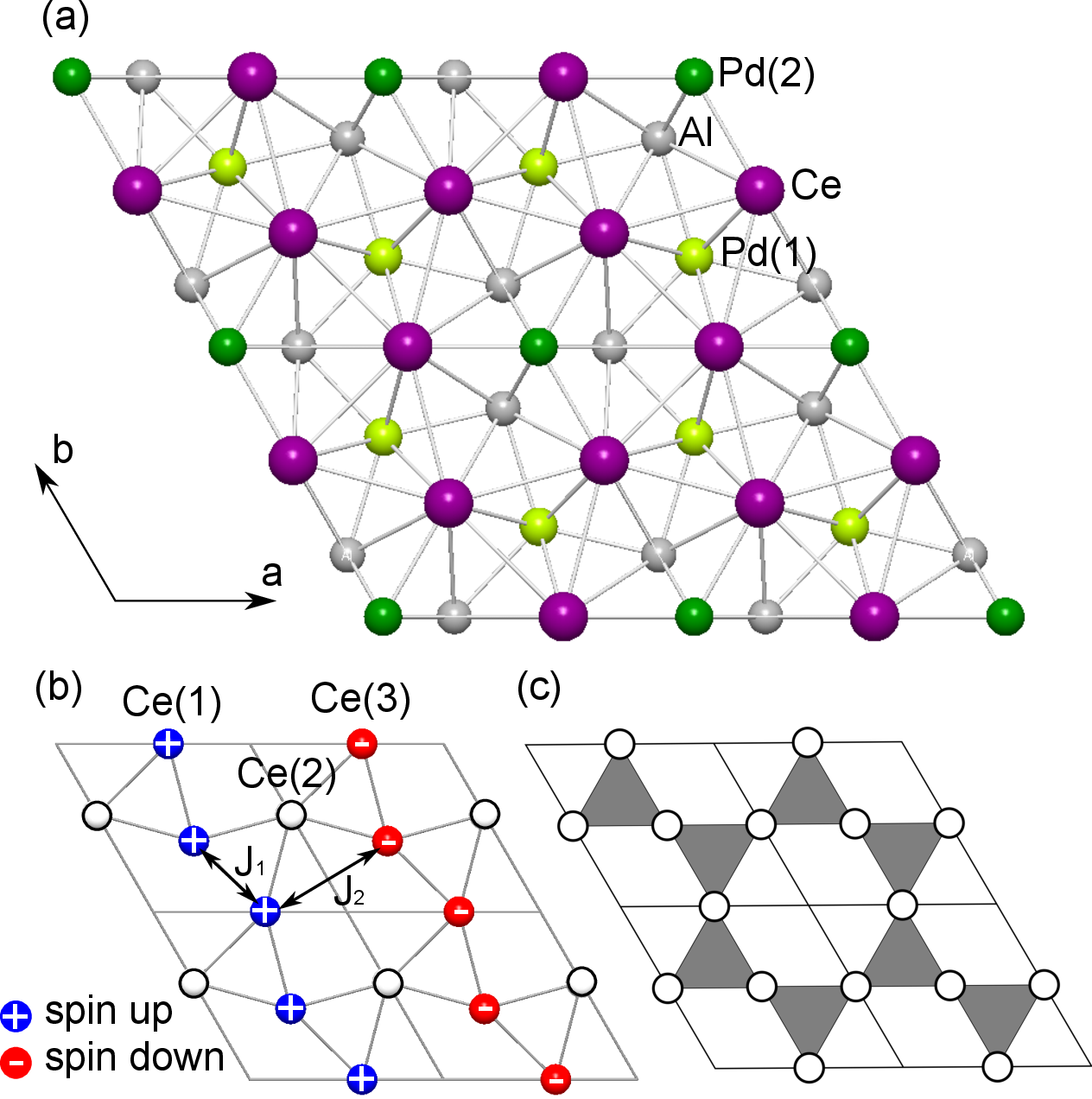}  \end{center}
\caption{(color online). (a) View of the $ab$ plane of CePdAl. The Ce and Pd(2) atoms form a plane at  $z = 0$, the Al and Pd(1) atoms form a plane at $z= 1/2$.
(b) View of the $ab$ plane, showing one of the three symmetry-related  magnetic structures as derived from ref.~\onlinecite{Donni1996}. Only Ce atoms are shown for clarity.
(c) Kagom\'{e} lattice for comparison.
\label{struktur}}
\end{figure}

Geometric frustration is a possible means to access quantum critical behavior in magnets, with the possible appearance of spin li\-quids, e.g., in kagom\'{e} lattices.\cite{Lee2008} Consequently,  a more general (`global') phase diagram of quantum criticality was predicted.\cite{Si2006,Vojta2008,Coleman2010} However, QPTs induced by geometric frustration have been studied hitherto for insulating systems only. Hence it is of considerable interest to study quantum criticality in HF metals with partial frustration. First experiments were done recently on Ce$_2$Pt$_2$Pb, a realization of the Shastry-Sutherland model,\cite{Kim2011} and  partially frustrated YbAgGe.\cite{Budko2004} In the latter system and in Yb$_2$Pt$_2$Pb (Ref. \onlinecite{Kim2013}) magnetic-field induced QPTs were studied.

In this paper we investigate the quantum critical behavior of partially frustrated \mbox{CePdAl}, a HF system with a distorted kagom\'{e} lattice, see Fig.~\ref{struktur} (a). In its hexagonal ZrNiAl-type crystal structure, the magnetic Ce ions form a network of equilateral corner-sharing triangles in the $ab$ plane.\cite{Oyamada1996,Donni1996} The compound exhibits a strong magnetic anisotropy with the susceptibility ratio $\chi_c/\chi_{ab} \approx 14$ which is attributed to crystalline-electric-field and exchange anisotropies.\cite{Isikawa1996} CePdAl exhibits Kondo-lattice properties as evidenced by an increase of the electrical resistivity towards low temperature yielding  a Kondo temperature $T_{\rm K} \approx 5$~K,\cite{Goto2002} and orders antiferromagnetically below $T_{\rm N} = 2.7\,\unit{K}$. A partially ordered magnetic state of CePdAl was revealed by neutron diffraction measurements on powder samples,\cite{Donni1996} with one third of the Ce moments not participating in the long-range order. The magnetic structure of CePdAl consists of three inequivalent Ce sites, with a magnetic ordering vector $\mathbf{Q} = (\frac{1}{2}\;0\;\tau)$ where $\tau \approx 0.35$ weakly depends on temperature.\cite{Prokevs2006} Within a single kagom\'{e}-like layer, ferromagnetic chains separated by non-ordered Ce(2) atoms are coupled antiferromagnetically (see Fig.~\ref{struktur} (b)). A recent thorough $^{27}$Al NMR study indicated that the partially ordered state is indeed stable down to at least $30$ mK (Ref.~\onlinecite{Oyamada2008}).

The nearest-neighbor (nn) and next-nearest neighbor (nnn) Ce-Ce distances within the basal plane are $d_{ab1} = 3.73\,\unit{\AA}$  and $d_{ab2} = 5.25\,\unit{\AA}$, respectively.\cite{Donni1996,Pearsons2012} The interplane nn Ce-Ce distance is $d_{\perp} = 4.24\,\unit{\AA}$.\cite{Donni1996} Neglecting the interplane interactions, the magnetic structure has been described in terms of a purely $2$D model with nn ferromagnetic ($J_1$) and nnn AF ($J_2$) interactions within the $ab$ plane,\cite{Nunez-Regueiro1997} see Fig.~\ref{struktur} (b). The competition between Kondo and RKKY interactions was modeled in this approach by a $T$-dependent on-site parameter $\Delta_i (T)$, and was studied in detail by variational Monte Carlo simulations.\cite{Motome2010} Unfortunately, a model incorporating the interplane  coupling $J_{\perp}$ to be compared with the experimentally determined magnetic structure does not exist up to now. In the view of the incommensurate $z$ component $\tau$, even long-range interactions might have to be taken into account. Previous work on CePdAl  showed that $T_{\rm N}$ can be suppressed  by hydrostatic pressure \cite{Goto2002} or partial substitution of Pd by Ni,\cite{Isikawa2000,Fritsch2013}  suggesting the possibility for a QCP.

\section{Experimental Details}
\begin{figure}\begin{center}
\includegraphics[clip,width=\columnwidth]{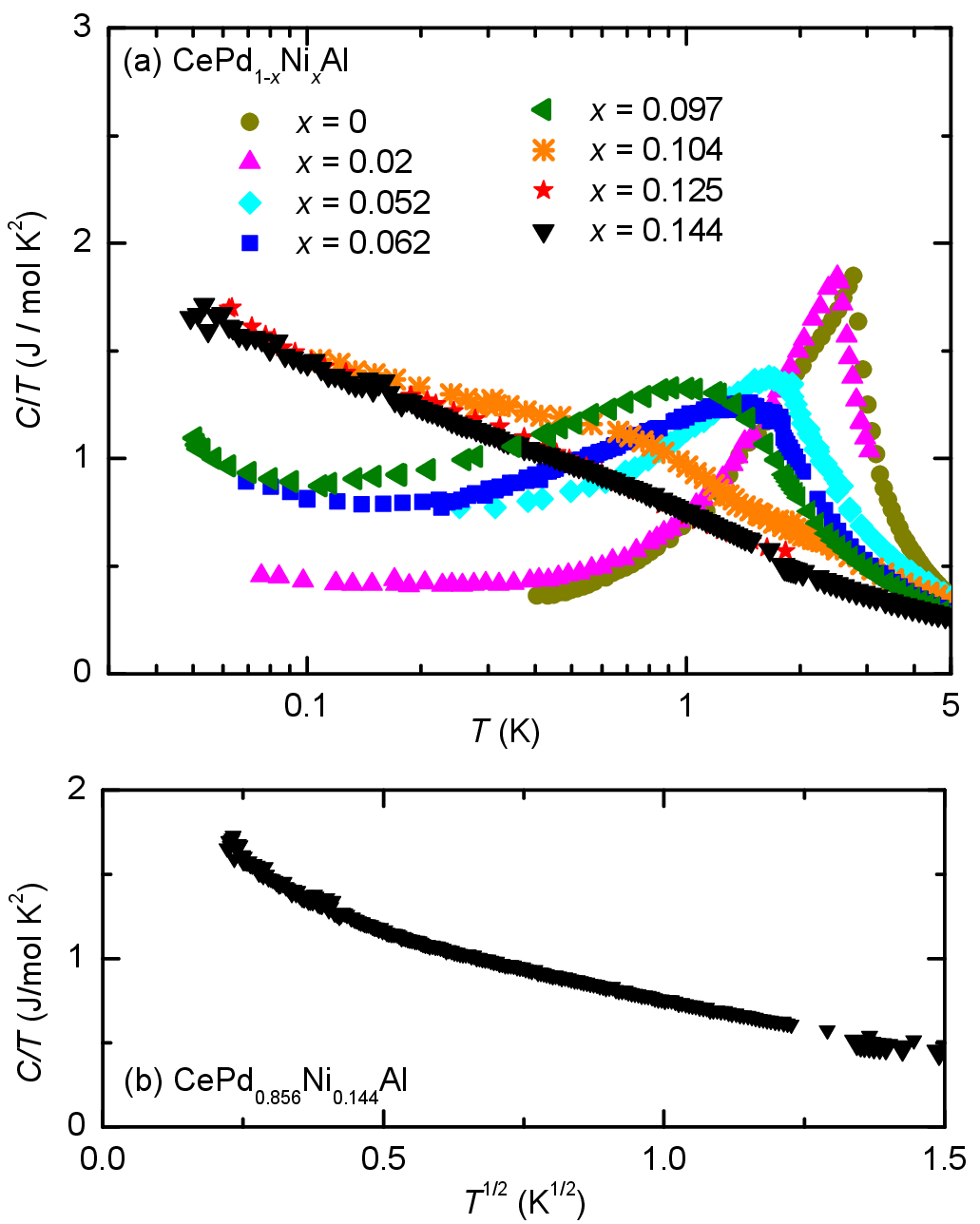}\end{center}
\caption{(color online). (a) Specific heat $C$ of CePd$_{1-x}$Ni$_{x}$Al plotted as $C/T$ versus $\log\,T$.
(b)~Specific heat $C$ of CePd$_{0.856}$Ni$_{0.144}$Al plotted as $C/T$ versus $T^{\frac{1}{2}}$. \label{CTvsT}}
\end{figure}

Polycrystalline samples of CePd$_{1-x}$Ni$_x$Al were prepared by arc-melting appropriate amounts of the pure elements Ce (Ref. \onlinecite{ames}), Pd(99.95), Ni(99.95), Al(99.999) under argon atmosphere with titanium gettering. To achieve homogeneity, the samples were remelted several times. The total weight loss after preparation did not exceed $0.5\%$. The samples were investigated in the as-cast state since annealing may cause a structural change.\cite{Gribanov2006} They were characterized by powder x-ray diffraction, revealing the single-phase \mbox{ZrNiAl} structure $(P\overline{6}2m)$ of the parent compounds. Atom absorption spectroscopy  was used to determine the actual Ni concentrations $x$ that are quoted throughout this paper.
The lattice constants $a$ and $c$ and the unit-cell volume $V$ approximately follow Vegard's law in the concentration range investigated ($x < 0.15$).  Specific-heat measurements  were performed in the temperature range $0.05 \leq T \lesssim 2.5$~K using the standard heat-pulse technique. A Physical Properties Measurement System (PPMS, Quantum Design) was used to obtain data at higher temperatures for some samples. The dc magnetic susceptibility $\chi$ was measured at $0.1$~T in the zero-field-cooled field-heated mode in a vibrating sample magnetometer (VSM, Oxford Instruments). A sample-dependent residual background contribution $\chi_0 \approx 2  \cdot 10^{-4}\,\mu_B/$T\,f.u. independent of $T$  was subtracted from the data. Below $20$~K, $\chi_0$ corresponds  to $< 1\%$ of the total susceptibility $\chi$.

\section{Results}
The specific heat $C$ is shown as $C/T$ vs. $\log T$ in Fig.~\ref{CTvsT} (a). The pure CePdAl compound exhibits a sharp anomaly at the N\'{e}el temperature $T_{\rm N} = 2.7$~K in agreement with literature data.\cite{Schank1994} The anomaly broadens and  moves to lower $T$ with increasing Ni content $x$ indicating a suppression of the antiferromagnetic (AF) transition. For $x = 0.144$, the $C/T$ vs. $\log T$ data follow a straight line in Fig.~\ref{CTvsT} (a) over almost two decades of temperature in the range $0.05 \leq T \leq 3$~K, i.e., $C/T = a \log (T_0/T)$, with $a=0.705$~J/mol K$^2$ and $T_0=11.7$~K. The non-Fermi-liquid behavior in the form of a logarithmic divergence of $C/T$ versus $T$ extends  over nearly two orders of magnitude in $T$ ($0.05 - 3$ K). The HMM model \cite{Hertz1976,Millis1993,Moriya1995}  predicts for $2$D AF fluctuations a logarithmic dependence of $C/T$ near the QCP as observed for CePd$_{1-x}$Ni$_x$Al, while the HMM prediction for $3$D antiferromagnets, $C/T \propto \gamma - a\sqrt{T}$ for $T \rightarrow 0$, is clearly not compatible with our data, as can be seen from the plot of $C/T$ vs  $\sqrt{T}$  in Fig~\ref{CTvsT} (b).

\begin{figure}
\begin{center}
\includegraphics[clip,width=\columnwidth]{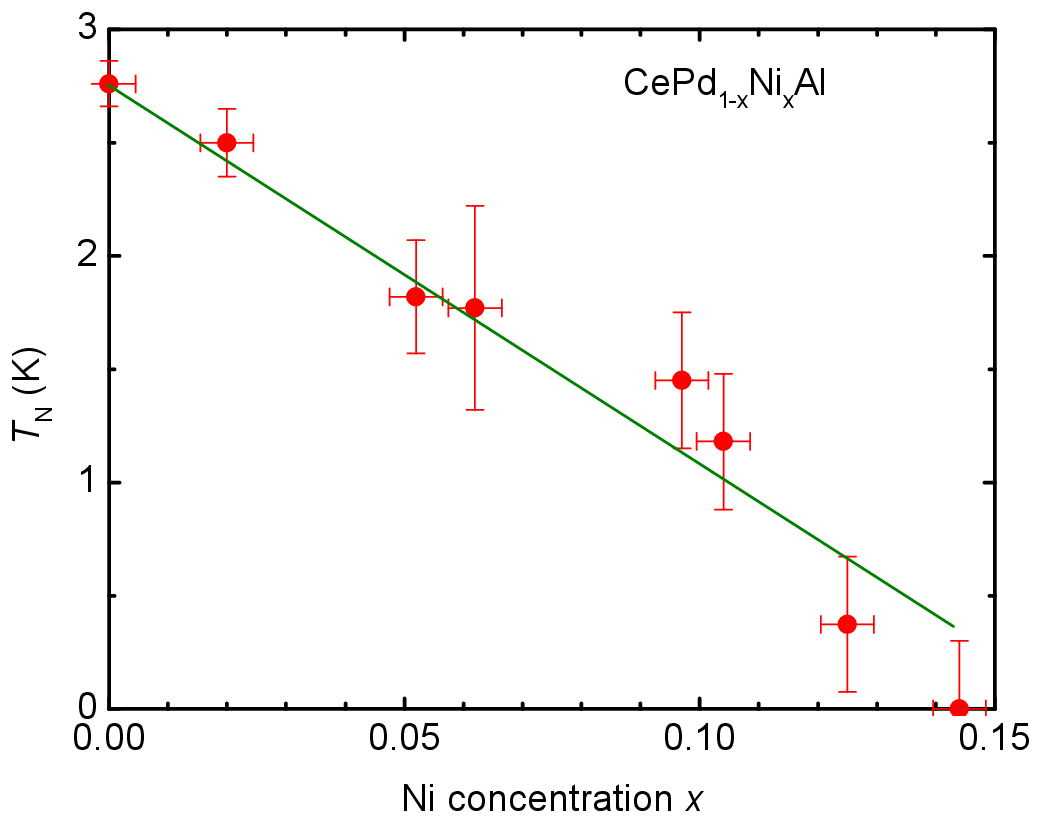} \end{center}
\caption{(color online). N\'{e}el temperature $T_{N}$ vs. Ni concentration $x$ of CePd$_{1-x}$Ni$_{x}$Al. \label{px}}
\end{figure}
We use the specific-heat data to obtain $T_{\rm N}$  as the temperature where $C(T)$ is a maximum for low $x$. For larger Ni concentrations, however, the small specific-heat  anomaly  associated with the onset of magnetic order resides on a large $T$-dependent background $C_{\rm bg}$, similar to the behavior of CeCu$_{6-x}$Au$_x$ close to the
QCP.\cite{Lohneysen1996}  Hence we use $C(T)$  of the allegedly quantum-critical sample with $x = 0.144$ as 'background' that is subtracted from the data of the samples close to $x_c$, and take $T_{\rm N}$  as the temperature of the maximum of the resulting $\Delta C(T)$ curve. Of course, we cannot pin down the critical concentration  with absolute accuracy. If the critical concentration would be slightly above $0.144$, we would estimate for $x = 0.144$ a maximum possible broad anomaly at $T_{\rm N}$ to $\approx 0.15\,\unit{K}$ for this sample. The resulting $T_{\rm N}(x)$ is shown in Fig.~\ref{px}. Up to $x = 0.1$, $T_{\rm N}(x)$ nicely follows a straight line as expected for the $2$D~AF QCP in the HMM model. The deviations towards the QCP, possibly signaling a restauration of three-dimensionality, will be discussed below.

One might want to compare  $T_{\rm N}(x)$ with the pressure dependence of $T_{\rm N}$ of pure CePdAl via the changes of the unit-cell volume  by external hydrostatic pressure $p$ or by chemical pressure exerted by the smaller Ni atoms. While $V(x)$ is directly accessible from our x-ray diffraction data, for $V(p)$ one needs the compressibility which can be roughly estimated from the bulk moduli $B_V$ of the pure constituents as described in Ref.~\onlinecite{Grimvall1999}. Apart from this uncertainty, $T_{\rm N} (p)$ data of different authors,\cite{Tang1996,Goto2002,Prokevs2006} while agreeing on a reduction of $T_{\rm N}$ with $p$, differ by a factor of $\approx 2$. We therefore cannot conclude whether the suppression of $T_{\rm N}$ with $x$ is solely due to a volume effect  (if the strong $T_{\rm N}(p)$ dependence of Goto {\it et al.} (Ref. \onlinecite{Goto2002}) is valid) or is mediated partly by a change of the electronic structure induced by Ni substitution.

\begin{figure}\begin{center}
\includegraphics[clip,width=0.9\columnwidth]{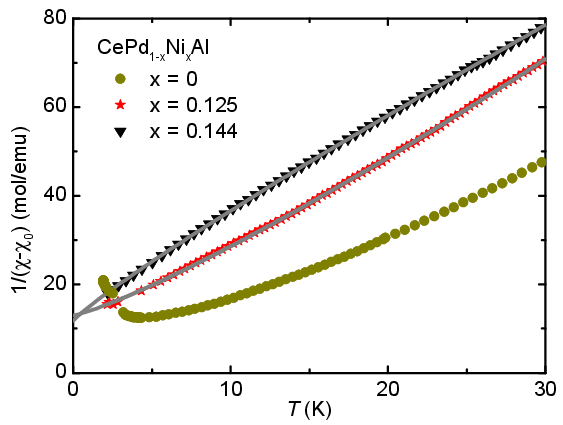}\end{center}
\caption{(color online). Inverse susceptibility $(\chi - \chi_0)^{-1}$ of CePd$_{1-x}$Ni$_x$Al versus temperature $T$ for $x = 0$, $0.125$ and $0.144$ in the low temperature region below $30$~K, where a distinct difference in curvature emerges when approaching the QCP. A sample-dependent residual background contribution $\chi_0$ independent of $T$ (see text) was subtracted from the data. The   lines show a fit with a power law $1/(\chi - \chi_0) \propto T^{\alpha} + \Theta^{\alpha} $ with $\alpha = 1.2$ for $x = 0.125$ and $\alpha = 0.9$ for $x = 0.144$.   \label{invchi}}
\end{figure}

Fig.~\ref{invchi} shows the low-temperature data of the inverse susceptibility $1/\left(\chi-\chi_0\right)$ vs. $T$ for $x = 0$, $0.125$ and $0.144$ (Ref.~\onlinecite{Fritsch2013}). The data reveal that, at low $T$, a marked difference in the $T$ dependence develops. While $\chi^{-1} (T)$ varies superlinearly for $x = 0$ approaching $T_{\rm N}$ from $T > T_{\rm N}$, the data for $x = 0.144$ can be described with an exponent $\alpha \approx 0.9$. This might point to an anomalous scaling exponent $\alpha < 1$ as observed for CeCu$_{6-x}$Au$_x$.\cite{Schroder1998,Schroder2000} Note that this sublinear exponent is only observed close to the QCP. Already for $x = 0.125$ the susceptibility exponent is $\alpha = 1.2$ signalling the finite $T_{\rm N} = 0.374$~K. Indeed, a sublinear exponent is expected in the model of local quantum criticality.\cite{Coleman2001}

\section{Discussion}
In geometrically frustrated {\em insulating} magnetic systems with stable moments, frustration arises from (near) cancelation of competing exchange interactions at the site of  a given moment. In this case the frustration parameter ist defined as $f = \left|\Theta_{\rm CW}\right|/T_{\rm N}$, where $\Theta_{\rm CW}$ is the Curie-Weiss temperature derived from the susceptibility in the paramagnetic regime. Using this definition, it could be shown for the poorly metallic $RX$Cu$_4$ system with $R = $ rare-earth element and $X =$ In or Cd that there is an approximate inverse relation between $f$ and the electrical conductivity $\sigma$, i.e., higher $\sigma$ leads to lower $f$.\cite{Fritsch2006b} This might explain why relatively few frustrated {\em metallic} systems have been investigated so far. In CePd$_{1-x}$Ni$_x$Al $\left|\Theta_{\rm CW}\right|$ increases slightly with $x$ (by a factor of $\approx 2$), while $T_{\rm N}$ decreases all the way to $T_{\rm N} =0$ between $x =0$ and $x=0.144$ (Ref.~\onlinecite{Fritsch2013}). The apparent increase of the ratio $\left|\Theta_{\rm CW}\right|/T_{\rm N}$ with $x$ thus is mostly due to a reduction of $T_{\rm N}$ and
rather reflects the competition between Kondo effect and RKKY interactions than an increased geometric frustration.  In fact, a similar apparent increase of $\left|\Theta_{\rm CW}\right|/T_{\rm N}$ can be inferred from data for other HF systems approaching quantum criticality that do not exhibit  geometric frustration.

The disorder introduced by the partial replacement of Pd by Ni leads for low Ni concentration $x \ll 1$, i. e., to first order in $x$, to Ce atoms having two different local environments and thus possibly two different Kondo temperatures $T_{\normalfont K}$. This is quite different from the strong disorder  arising from the wide $T_{\normalfont K}$ distribution of  local magnetic moments stemming from the random distribution of dopant atoms in semiconductors close to the metal-insulator transition \cite{Lakner1994}, and also observed in some U alloys such as UCu$_{5-x}$Pd$_x$ (\cite{Miranda1997} and refs. therein). In fact, a comparison of NMR and $\mu$SR experiments for UCu$_5$ doped with Pd
and CeCu$_6$ doped with Au \cite{Bernal1996}
showed decisive differences in the type and degree of disorder, with the
Ce system being much less disordered compared to the U system. Furthermore pressure measurements on CeCu$_{5.5}$Au$_{0.5}$ have shown that disorder has no decisive role in this system \cite{Hamann2013}.
The wide $T_{\normalfont K}$ distribution leads to a resistivity $\rho$ that {\em increases} with decreasing $T$ due to the presence of magnetic moments with $T_{\normalfont K} < T$ and, for certain distributions, to a logarithmic $T$ dependence of $C/T$ \cite{Miranda1997}. However, preliminary measurements of $\rho$ for CePd$_{1-x}$Ni$_x$Al with $x = 0.144$ show a {\em decrease} of $\rho$ with decreasing $T$ below $5$~K. We can therefore dismiss the distribution of $T_{\normalfont K}$ as a primary source of our experimental observations.

\begin{figure}
\begin{center}
\includegraphics[clip,width=\columnwidth]{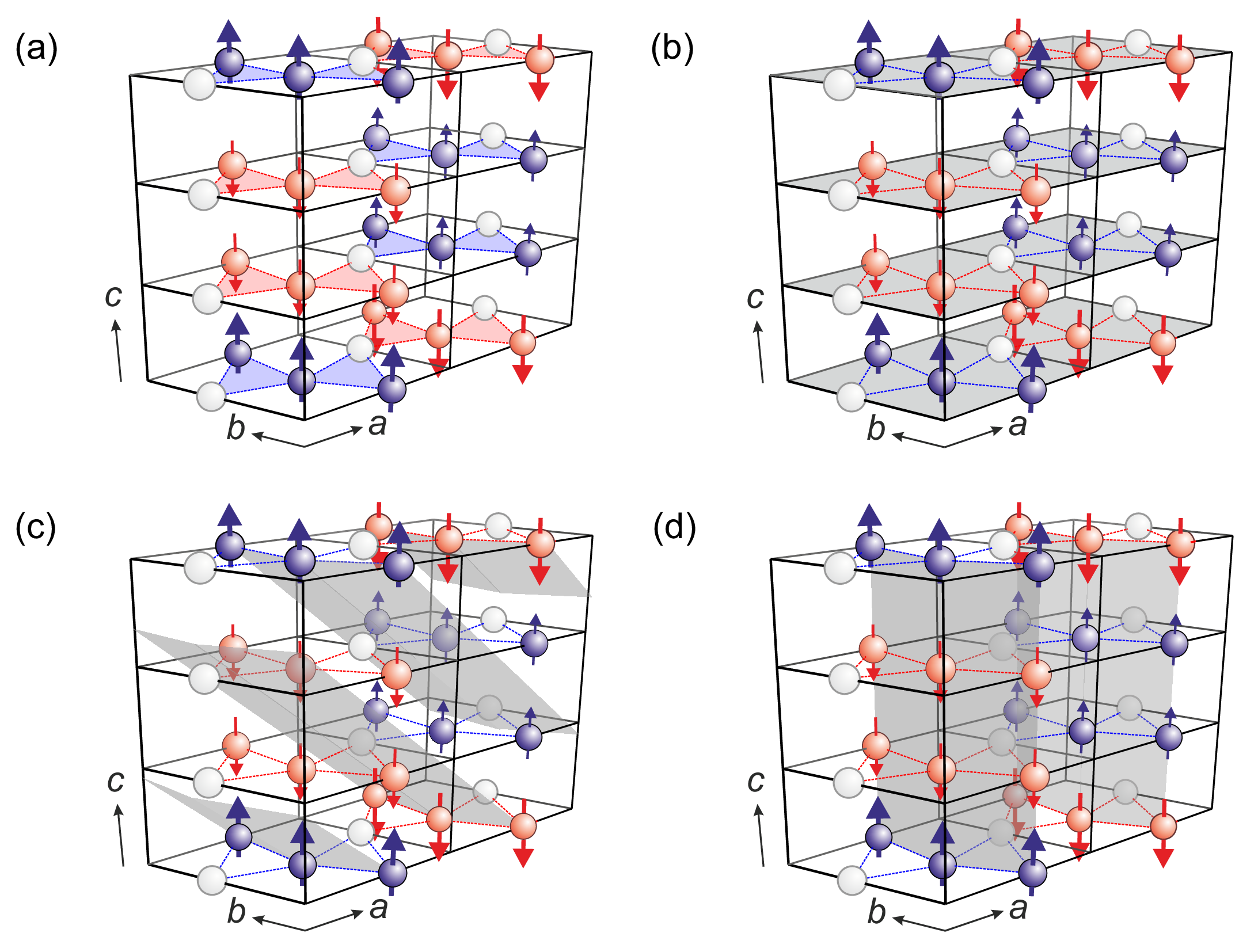}
\end{center}
\caption{(color online). (a) Three-dimensional magnetic structure of \mbox{CePdAl} (only Ce atoms are shown) illustrating the ferromagnetic chains in the $ab$-plane along the $b$-axis. (b) Kagom\'{e}-like $ab$ planes. (c)  $2$D ferromagnetic planes perpendicular to the magnetic ordering wave vector $\mathbf{Q}$. (d) $2$D antiferromagnetic planes spanned by the ferromagnetic chains in the $ab$ plane and the perpendicular $c$ direction and separated by frustrated planes of Ce(2) atoms.  \label{planes}}\end{figure}
To address the role of frustration concerning the QCP in \mbox{CePdAl,} we first discuss this issue in terms of the HMM model. While the magnetic  structure of CePdAl is three-dimensional in nature with the ordering wave vector $\mathbf{Q} = (\frac{1}{2}\;0\;\tau)$, the specific heat $C/T \propto \log (T_0/T)$ for $x \approx x_c$ and the linear $T_{\rm N} (x)$ dependence are indeed in accord with $2$D AF quantum fluctuations in the HMM scenario. For a single kagom\'{e}-like $ab$ plane, the magnetic structure can be described by ferromagnetic chains, as mentioned above, that couple only weakly antiferromagnetically by the nnn interaction $J_2$, because of the frustration of the nn interactions $J_1$ at the Ce$(2)$ sites.\cite{Nunez-Regueiro1997,Motome2010}

With recurrence to the three-dimensional magnetic structure of CePdAl \cite{Donni1996,Prokevs2006} schematically shown in Fig.~\ref{planes} (a) assuming for clarity $\tau = 1/3$, we can try to identify possible $2$D AF planes: (1) the kagom\'{e}-like $ab$ planes (Fig.~\ref{planes} (b)) cannot be taken as origin of the two-dimensional antiferromagnetic fluctuations because the interlayer exchange $J_{\perp}$ between Ce atoms  is expected to be stronger than $J_2$. (2) Taking $J_{\perp}$ into account, the magnetic order can be viewed (Fig.~\ref{planes} (c)) as a stacking of ferromagnetic chains in the $ab$ plane staggered along the $c$-direction and  interrupted by the frustrated Ce(2) moments, thus leading to $2$D  ferromagnetic planes oriented roughly perpendicularly to  $\mathbf{Q}$. However, $2$D ferromagnetic fluctuations will in the HMM model give rise to a specific heat $C/T \propto T^{-\frac{1}{3}}$, which is not supported by our data. (3) A triple-$\mathbf{Q}$ structure is in principle compatible with the neutron-scattering  data.\cite{Donni1996}  However,  the magnetic component $\tau$ along the $c$ direction is, in fact, incommensurate and varies slightly with $T$ ($\tau = 0.3535 - 0.3585$ between $0.35$ and $2.7$~K).\cite{Prokevs2006}  This $T$-dependent incommensurability is not easy to explain within a triple-$\mathbf{Q}$ structure expected to lock to $\tau = \frac{1}{3}$. Furthermore, $2$D magnetic structures are not formed in the triple-Q order of CePdAl. (4) Planes spanned by the ferromagnetic chains in the $ab$ plane and the perpendicular $c$ direction (Fig.~\ref{planes} (d)) are indeed antiferromagnetic. Because of the distortion of the kagom\'{e} planes they are corrugated and they are separated by frustrated Ce$(2)$ moments and could give rise to  $2$D AF fluctuations. Hence of the possibilities listed, only the last one presents a viable explanation for the $C/T \propto \log \left(T_0/T\right)$ behavior.

In the classical case of an anisotropic $3$D system geometric frustration might give way to strictly $2$D order, while in a quantum system zero-point fluctuations might restore the original dimensionality.\cite{Schmalian2008} This, however, is not always the case. For example, it was shown in the context of a Bose-Einstein condensation of magnetic triplets in BaCuSi$_2$O$_6$ \cite{Sebastian2006} that geometrical frustration yields a reduction of the spatial dimensionality even close to the QCP.\cite{Batista2007,Schmalian2008}
The partial magnetic frustration in CePdAl appears to reduce the dimensionality by forming frustrated planes from one third of the Ce moments, decoupling the antiferromagnetic planes of the other Ce moments as argued above. CePdAl might be a unique example where frustration provides a rationale of  how a system might find its way from $3$D magnetic ordering to $2$D criticality, a {\em microscopic} link that is missing for CeCu$_{6-x}$Au$_x$, where the origin of  $2$D fluctuations remains unclear. Thus CePdAl might also be a very promising candidate to test the recent predictions for the $2$D-$3$D crossover in terms of the critical quasiparticle theory.\cite{Abrahams2012} The deviations of $T_{\mathrm{N}}(x)$ from a linear $x$ dependence close to $x_c$ (cf. Fig.~\ref{px}) are compatible with such a crossover.

While we have discussed the possibility of $2$D fluctuations confined to AF planes, the  intermediate frustrated planes of Ce(2) atoms certainly deserve attention in their own right as well. Here we discuss the  perspective of a $2$D spin liquid of these Ce(2) atoms that form a rectangular lattice (lattice constants $d_{\perp} =  4.24\,\unit{\AA}$ and $d_{Ce(2)} = 7.2170\,\unit{\AA}$ perpendicular and parallel to the kagom\`{e} planes). Note that these Ce(2) planes are not corrugated.
There are several arguments supporting the fascinating possibility of a $2D$ spin liquid in CePdAl. First of all, no further phase transition below $T_{\mathrm N}$ occurs in zero field down to $30\,\unit{mK}$ as shown by detailed NMR experiments \cite{Oyamada2008}. Furthermore these measurements suggest a gapless excitation spectrum for the paramagnetic third of Ce-ions, similar to the findings in the proposed spin liquid ZnCu$_3$(OH)$_6$Cl$_{2}$ \cite{Helton2007}. Of course,  the Kondo effect acting on the Ce(2) atoms would be a possibility. However, the resistivity $\rho(T)$  of pure CePdAl decreases monotonically toward low $T$.  Furthermore, $\rho(B)$ at low T increases in an external magnetic field up to $B = 3.4\,\unit{T}$ \cite{Goto2002}, where a transition to a phase involving the hitherto not ordered Ce(2) moments occurs \cite{Prokevs2006}. On the theoretical side, the $T =0$ ground state of a non-ordered Kondo lattice in general exhibits Kondo-screened moments, i.e.,  is a Fermi li\-quid regardless of the magnitude of the onsite Kondo coupling $J_K$. Only when frustrating intersite exchange interactions are included, it is possible to find a breakdown of Kondo screening below a finite Kondo coupling $J_{K_c}$. In this case a spin liquid state for $J_K < J_{K_c}$ may evolve even for frustrated $3D$ systems and, a fortiori, for frustrated $2D$ systems such as the Ce(2) planes in CePdAl \cite{Senthil2004}.

Returning to the QCP of CePd$_{1-x}$Ni$_x$Al, it is important to point out that in the present treatments of the HMM model \cite{Hertz1976,Millis1993,Moriya1995} quantum fluctuations arising from geometric frustration  are not included. Hence other scenarios are conceivable to explain the quantum critical behavior of Ni-subsituted CePdAl. Indeed, the susceptibility exponent $\alpha < 1$ for $x = 0.144$ could possibly be due to anomalous scaling.
Detailed experiments including inelastic neutron scattering to determine scaling relations between the energy, temperature and field dependencies of quantum critical fluctuations as done, e.g., for CeCu$_{6-x}$Au$_x$ \cite{Schroder2000}, are necessary to narrow down the origin of the unusual quantum critical behavior of CePdAl. As a further interesting point, CePdAl offers the possibility of investigating the Kondo effect  of geometrically frustrated metals in detail, which may lead to a better understanding of the `global phase diagram' \cite{Si2006,Vojta2008} of quantum criticality in metallic systems.

\section{Conclusion}
We have investigated systematically the effect of chemical pressure on the specific heat of CePd$_{1-x}$Ni$_x$Al alloys for  Ni concentrations ranging up to $x = 0.144$, where the magnetic order is completely suppressed, giving rise to a quantum critical point. Here the observed $T \log \left(T_0/T\right)$ dependence of the specific heat suggests, if interpreted within the HMM scenario of conventional quantum criticality, the evolution of $2$D magnetic fluctuations. The partial geometric frustration of Ce moments in this system may  be instrumental in establishing the unusual $2$D-like character of quantum criticality, despite the fact that the magnetic order itself is $3$D. On the other hand, the partial frustration leading to an only partially ordered magnetic ground state might lead to novel types of excitations close to the QCP not contained in the HMM model. In this respect, our results present important clues on how a two-dimensional spin-liquid-like collective of frustrated moments immersed into long-range antiferromagnetic order reacts to the suppression of magnetic order to zero temperature near quantum criticality.

\begin{acknowledgments}
We thank J. Schmalian and O. Stockert for fruitful discussions and S. Drobnik for help with some measurements. This work was supported by the Deutsche Forschungsgemeinschaft through FOR 960. H.v.L. thanks the National Science Foundation for partial support of this work under Grant No. PHYS-1066293, and the Aspen Center for Physics for its hospitality.
\end{acknowledgments}


\begin{thebibliography}{52}%
\makeatletter
\providecommand \@ifxundefined [1]{%
 \@ifx{#1\undefined}
}%
\providecommand \@ifnum [1]{%
 \ifnum #1\expandafter \@firstoftwo
 \else \expandafter \@secondoftwo
 \fi
}%
\providecommand \@ifx [1]{%
 \ifx #1\expandafter \@firstoftwo
 \else \expandafter \@secondoftwo
 \fi
}%
\providecommand \natexlab [1]{#1}%
\providecommand \enquote  [1]{``#1''}%
\providecommand \bibnamefont  [1]{#1}%
\providecommand \bibfnamefont [1]{#1}%
\providecommand \citenamefont [1]{#1}%
\providecommand \href@noop [0]{\@secondoftwo}%
\providecommand \href [0]{\begingroup \@sanitize@url \@href}%
\providecommand \@href[1]{\@@startlink{#1}\@@href}%
\providecommand \@@href[1]{\endgroup#1\@@endlink}%
\providecommand \@sanitize@url [0]{\catcode `\\12\catcode `\$12\catcode
  `\&12\catcode `\#12\catcode `\^12\catcode `\_12\catcode `\%12\relax}%
\providecommand \@@startlink[1]{}%
\providecommand \@@endlink[0]{}%
\providecommand \url  [0]{\begingroup\@sanitize@url \@url }%
\providecommand \@url [1]{\endgroup\@href {#1}{\urlprefix }}%
\providecommand \urlprefix  [0]{URL }%
\providecommand \Eprint [0]{\href }%
\providecommand \doibase [0]{http://dx.doi.org/}%
\providecommand \selectlanguage [0]{\@gobble}%
\providecommand \bibinfo  [0]{\@secondoftwo}%
\providecommand \bibfield  [0]{\@secondoftwo}%
\providecommand \translation [1]{[#1]}%
\providecommand \BibitemOpen [0]{}%
\providecommand \bibitemStop [0]{}%
\providecommand \bibitemNoStop [0]{.\EOS\space}%
\providecommand \EOS [0]{\spacefactor3000\relax}%
\providecommand \BibitemShut  [1]{\csname bibitem#1\endcsname}%
\let\auto@bib@innerbib\@empty
\bibitem [{\citenamefont {Sachdev}\ and\ \citenamefont
  {Keimer}(2011)}]{Sachdev2011}%
  \BibitemOpen
  \bibfield  {author} {\bibinfo {author} {\bibfnamefont {S.}~\bibnamefont
  {Sachdev}}\ and\ \bibinfo {author} {\bibfnamefont {B.}~\bibnamefont
  {Keimer}},\ }\href@noop {} {\bibfield  {journal} {\bibinfo  {journal}
  {Physics Today}\ }\textbf {\bibinfo {volume} {64}},\ \bibinfo {pages} {29}
  (\bibinfo {year} {2011})}\BibitemShut {NoStop}%
\bibitem [{\citenamefont {Gegenwart}\ \emph {et~al.}(2008)\citenamefont
  {Gegenwart}, \citenamefont {Si},\ and\ \citenamefont
  {Steglich}}]{Gegenwart2008}%
  \BibitemOpen
  \bibfield  {author} {\bibinfo {author} {\bibfnamefont {P.}~\bibnamefont
  {Gegenwart}}, \bibinfo {author} {\bibfnamefont {Q.}~\bibnamefont {Si}}, \
  and\ \bibinfo {author} {\bibfnamefont {F.}~\bibnamefont {Steglich}},\
  }\href@noop {} {\bibfield  {journal} {\bibinfo  {journal} {Nature Physics}\
  }\textbf {\bibinfo {volume} {4}},\ \bibinfo {pages} {186} (\bibinfo {year}
  {2008})}\BibitemShut {NoStop}%
\bibitem [{\citenamefont {Greiner}\ \emph {et~al.}(2002)\citenamefont
  {Greiner}, \citenamefont {Mandel}, \citenamefont {Esslinger}, \citenamefont
  {H{\"a}nsch},\ and\ \citenamefont {Bloch}}]{Greiner2002}%
  \BibitemOpen
  \bibfield  {author} {\bibinfo {author} {\bibfnamefont {M.}~\bibnamefont
  {Greiner}}, \bibinfo {author} {\bibfnamefont {O.}~\bibnamefont {Mandel}},
  \bibinfo {author} {\bibfnamefont {T.}~\bibnamefont {Esslinger}}, \bibinfo
  {author} {\bibfnamefont {T.~W.}\ \bibnamefont {H{\"a}nsch}}, \ and\ \bibinfo
  {author} {\bibfnamefont {I.}~\bibnamefont {Bloch}},\ }\href@noop {}
  {\bibfield  {journal} {\bibinfo  {journal} {Nature}\ }\textbf {\bibinfo
  {volume} {415}},\ \bibinfo {pages} {39} (\bibinfo {year} {2002})}\BibitemShut
  {NoStop}%
\bibitem [{\citenamefont {Doniach}(1977)}]{Doniach1977}%
  \BibitemOpen
  \bibfield  {author} {\bibinfo {author} {\bibfnamefont {S.}~\bibnamefont
  {Doniach}},\ }\href@noop {} {\bibfield  {journal} {\bibinfo  {journal}
  {Physica}\ }\textbf {\bibinfo {volume} {91B}},\ \bibinfo {pages} {231 }
  (\bibinfo {year} {1977})}\BibitemShut {NoStop}%
\bibitem [{\citenamefont {v.~L{\"o}hneysen}\ \emph {et~al.}(2007)\citenamefont
  {v.~L{\"o}hneysen}, \citenamefont {Rosch}, \citenamefont {Vojta},\ and\
  \citenamefont {W{\"o}lfle}}]{Lohneysen2007}%
  \BibitemOpen
  \bibfield  {author} {\bibinfo {author} {\bibfnamefont {H.}~\bibnamefont
  {v.~L{\"o}hneysen}}, \bibinfo {author} {\bibfnamefont {A.}~\bibnamefont
  {Rosch}}, \bibinfo {author} {\bibfnamefont {M.}~\bibnamefont {Vojta}}, \ and\
  \bibinfo {author} {\bibfnamefont {P.}~\bibnamefont {W{\"o}lfle}},\
  }\href@noop {} {\bibfield  {journal} {\bibinfo  {journal} {Rev. Mod. Phys.}\
  }\textbf {\bibinfo {volume} {79}},\ \bibinfo {pages} {1015} (\bibinfo {year}
  {2007})}\BibitemShut {NoStop}%
\bibitem [{\citenamefont {Stockert}\ \emph {et~al.}(1998)\citenamefont
  {Stockert}, \citenamefont {{v. L{\"o}hneysen}}, \citenamefont {Rosch},
  \citenamefont {Pyka},\ and\ \citenamefont {Loewenhaupt}}]{Stockert1998}%
  \BibitemOpen
  \bibfield  {author} {\bibinfo {author} {\bibfnamefont {O.}~\bibnamefont
  {Stockert}}, \bibinfo {author} {\bibfnamefont {H.}~\bibnamefont {{v.
  L{\"o}hneysen}}}, \bibinfo {author} {\bibfnamefont {A.}~\bibnamefont
  {Rosch}}, \bibinfo {author} {\bibfnamefont {N.}~\bibnamefont {Pyka}}, \ and\
  \bibinfo {author} {\bibfnamefont {M.}~\bibnamefont {Loewenhaupt}},\
  }\href@noop {} {\bibfield  {journal} {\bibinfo  {journal} {Phys. Rev. Lett.}\
  }\textbf {\bibinfo {volume} {{80}}},\ \bibinfo {pages} {5627} (\bibinfo
  {year} {{1998}})}\BibitemShut {NoStop}%
\bibitem [{\citenamefont {{Schr{\"o}der}}\ \emph {et~al.}(1998)\citenamefont
  {{Schr{\"o}der}}, \citenamefont {Aeppli}, \citenamefont {Bucher},
  \citenamefont {Ramazashvili},\ and\ \citenamefont {Coleman}}]{Schroder1998}%
  \BibitemOpen
  \bibfield  {author} {\bibinfo {author} {\bibfnamefont {A.}~\bibnamefont
  {{Schr{\"o}der}}}, \bibinfo {author} {\bibfnamefont {G.}~\bibnamefont
  {Aeppli}}, \bibinfo {author} {\bibfnamefont {E.}~\bibnamefont {Bucher}},
  \bibinfo {author} {\bibfnamefont {R.}~\bibnamefont {Ramazashvili}}, \ and\
  \bibinfo {author} {\bibfnamefont {P.}~\bibnamefont {Coleman}},\ }\href@noop
  {} {\bibfield  {journal} {\bibinfo  {journal} {Phys. Rev. Lett.}\ }\textbf
  {\bibinfo {volume} {80}},\ \bibinfo {pages} {5623} (\bibinfo {year}
  {1998})}\BibitemShut {NoStop}%
\bibitem [{\citenamefont {Schr{\"o}der}\ \emph {et~al.}(2000)\citenamefont
  {Schr{\"o}der}, \citenamefont {Aeppli}, \citenamefont {Coldea}, \citenamefont
  {Adams}, \citenamefont {Stockert}, \citenamefont {L{\"o}hneysen},
  \citenamefont {Bucher}, \citenamefont {Ramazashvili},\ and\ \citenamefont
  {Coleman}}]{Schroder2000}%
  \BibitemOpen
  \bibfield  {author} {\bibinfo {author} {\bibfnamefont {A.}~\bibnamefont
  {Schr{\"o}der}}, \bibinfo {author} {\bibfnamefont {G.}~\bibnamefont
  {Aeppli}}, \bibinfo {author} {\bibfnamefont {R.}~\bibnamefont {Coldea}},
  \bibinfo {author} {\bibfnamefont {M.}~\bibnamefont {Adams}}, \bibinfo
  {author} {\bibfnamefont {O.}~\bibnamefont {Stockert}}, \bibinfo {author}
  {\bibfnamefont {H.}~\bibnamefont {L{\"o}hneysen}}, \bibinfo {author}
  {\bibfnamefont {E.}~\bibnamefont {Bucher}}, \bibinfo {author} {\bibfnamefont
  {R.}~\bibnamefont {Ramazashvili}}, \ and\ \bibinfo {author} {\bibfnamefont
  {P.}~\bibnamefont {Coleman}},\ }\href@noop {} {\bibfield  {journal} {\bibinfo
   {journal} {Nature}\ }\textbf {\bibinfo {volume} {407}},\ \bibinfo {pages}
  {351} (\bibinfo {year} {2000})}\BibitemShut {NoStop}%
\bibitem [{\citenamefont {Aronson}\ \emph {et~al.}(1995)\citenamefont
  {Aronson}, \citenamefont {Osborn}, \citenamefont {Robinson}, \citenamefont
  {Lynn}, \citenamefont {Chau}, \citenamefont {Seaman},\ and\ \citenamefont
  {Maple}}]{Aronson1995}%
  \BibitemOpen
  \bibfield  {author} {\bibinfo {author} {\bibfnamefont {M.~C.}\ \bibnamefont
  {Aronson}}, \bibinfo {author} {\bibfnamefont {R.}~\bibnamefont {Osborn}},
  \bibinfo {author} {\bibfnamefont {R.~A.}\ \bibnamefont {Robinson}}, \bibinfo
  {author} {\bibfnamefont {J.~W.}\ \bibnamefont {Lynn}}, \bibinfo {author}
  {\bibfnamefont {R.}~\bibnamefont {Chau}}, \bibinfo {author} {\bibfnamefont
  {C.~L.}\ \bibnamefont {Seaman}}, \ and\ \bibinfo {author} {\bibfnamefont
  {M.~B.}\ \bibnamefont {Maple}},\ }\href@noop {} {\bibfield  {journal}
  {\bibinfo  {journal} {Phys. Rev. Lett.}\ }\textbf {\bibinfo {volume} {75}},\
  \bibinfo {pages} {725} (\bibinfo {year} {1995})}\BibitemShut {NoStop}%
\bibitem [{\citenamefont {Coleman}\ \emph {et~al.}(2001)\citenamefont
  {Coleman}, \citenamefont {Pepin}, \citenamefont {Si},\ and\ \citenamefont
  {Ramazashvili}}]{Coleman2001}%
  \BibitemOpen
  \bibfield  {author} {\bibinfo {author} {\bibfnamefont {P.}~\bibnamefont
  {Coleman}}, \bibinfo {author} {\bibfnamefont {C.}~\bibnamefont {Pepin}},
  \bibinfo {author} {\bibfnamefont {Q.}~\bibnamefont {Si}}, \ and\ \bibinfo
  {author} {\bibfnamefont {R.}~\bibnamefont {Ramazashvili}},\ }\href@noop {}
  {\bibfield  {journal} {\bibinfo  {journal} {J. Phys.: Cond. Mat.}\ }\textbf
  {\bibinfo {volume} {13}},\ \bibinfo {pages} {R723} (\bibinfo {year}
  {2001})}\BibitemShut {NoStop}%
\bibitem [{\citenamefont {Si}\ \emph {et~al.}(2001)\citenamefont {Si},
  \citenamefont {Rabello}, \citenamefont {Ingersent},\ and\ \citenamefont
  {Smith}}]{Si2001}%
  \BibitemOpen
  \bibfield  {author} {\bibinfo {author} {\bibfnamefont {Q.}~\bibnamefont
  {Si}}, \bibinfo {author} {\bibfnamefont {S.}~\bibnamefont {Rabello}},
  \bibinfo {author} {\bibfnamefont {K.}~\bibnamefont {Ingersent}}, \ and\
  \bibinfo {author} {\bibfnamefont {J.}~\bibnamefont {Smith}},\ }\href@noop {}
  {\bibfield  {journal} {\bibinfo  {journal} {Nature}\ }\textbf {\bibinfo
  {volume} {{413}}},\ \bibinfo {pages} {804} (\bibinfo {year}
  {{2001}})}\BibitemShut {NoStop}%
\bibitem [{\citenamefont {Hertz}(1976)}]{Hertz1976}%
  \BibitemOpen
  \bibfield  {author} {\bibinfo {author} {\bibfnamefont {J.~A.}\ \bibnamefont
  {Hertz}},\ }\href@noop {} {\bibfield  {journal} {\bibinfo  {journal} {Phys.
  Rev. B}\ }\textbf {\bibinfo {volume} {14}},\ \bibinfo {pages} {1165}
  (\bibinfo {year} {1976})}\BibitemShut {NoStop}%
\bibitem [{\citenamefont {Millis}(1993)}]{Millis1993}%
  \BibitemOpen
  \bibfield  {author} {\bibinfo {author} {\bibfnamefont {A.~J.}\ \bibnamefont
  {Millis}},\ }\href@noop {} {\bibfield  {journal} {\bibinfo  {journal} {Phys.
  Rev. B}\ }\textbf {\bibinfo {volume} {48}},\ \bibinfo {pages} {7183}
  (\bibinfo {year} {1993})}\BibitemShut {NoStop}%
\bibitem [{\citenamefont {Moriya}\ and\ \citenamefont
  {Takimoto}(1995)}]{Moriya1995}%
  \BibitemOpen
  \bibfield  {author} {\bibinfo {author} {\bibfnamefont {T.}~\bibnamefont
  {Moriya}}\ and\ \bibinfo {author} {\bibfnamefont {T.}~\bibnamefont
  {Takimoto}},\ }\href@noop {} {\bibfield  {journal} {\bibinfo  {journal} {J.
  Phys. Soc. Jpn.}\ }\textbf {\bibinfo {volume} {64}},\ \bibinfo {pages} {960 }
  (\bibinfo {year} {1995})}\BibitemShut {NoStop}%
\bibitem [{\citenamefont {Paschen}\ \emph {et~al.}(2004)\citenamefont
  {Paschen}, \citenamefont {L{\"u}hmann}, \citenamefont {Wirth}, \citenamefont
  {Gegenwart}, \citenamefont {Trovarelli}, \citenamefont {Geibel},
  \citenamefont {Steglich}, \citenamefont {Coleman},\ and\ \citenamefont
  {Si}}]{Paschen2004}%
  \BibitemOpen
  \bibfield  {author} {\bibinfo {author} {\bibfnamefont {S.}~\bibnamefont
  {Paschen}}, \bibinfo {author} {\bibfnamefont {T.}~\bibnamefont
  {L{\"u}hmann}}, \bibinfo {author} {\bibfnamefont {S.}~\bibnamefont {Wirth}},
  \bibinfo {author} {\bibfnamefont {P.}~\bibnamefont {Gegenwart}}, \bibinfo
  {author} {\bibfnamefont {O.}~\bibnamefont {Trovarelli}}, \bibinfo {author}
  {\bibfnamefont {C.}~\bibnamefont {Geibel}}, \bibinfo {author} {\bibfnamefont
  {F.}~\bibnamefont {Steglich}}, \bibinfo {author} {\bibfnamefont
  {P.}~\bibnamefont {Coleman}}, \ and\ \bibinfo {author} {\bibfnamefont
  {Q.}~\bibnamefont {Si}},\ }\href@noop {} {\bibfield  {journal} {\bibinfo
  {journal} {Nature}\ }\textbf {\bibinfo {volume} {432}},\ \bibinfo {pages}
  {881} (\bibinfo {year} {2004})}\BibitemShut {NoStop}%
\bibitem [{\citenamefont {Garst}\ \emph {et~al.}(2008)\citenamefont {Garst},
  \citenamefont {Fritz}, \citenamefont {Rosch},\ and\ \citenamefont
  {Vojta}}]{Garst2008}%
  \BibitemOpen
  \bibfield  {author} {\bibinfo {author} {\bibfnamefont {M.}~\bibnamefont
  {Garst}}, \bibinfo {author} {\bibfnamefont {L.}~\bibnamefont {Fritz}},
  \bibinfo {author} {\bibfnamefont {A.}~\bibnamefont {Rosch}}, \ and\ \bibinfo
  {author} {\bibfnamefont {M.}~\bibnamefont {Vojta}},\ }\href@noop {}
  {\bibfield  {journal} {\bibinfo  {journal} {Phys. Rev. B}\ }\textbf {\bibinfo
  {volume} {78}},\ \bibinfo {pages} {235118} (\bibinfo {year}
  {2008})}\BibitemShut {NoStop}%
\bibitem [{\citenamefont {Custers}\ \emph {et~al.}(2012)\citenamefont
  {Custers}, \citenamefont {Lorenzer}, \citenamefont {M{\"u}ller},
  \citenamefont {Prokofiev}, \citenamefont {Sidorenko}, \citenamefont
  {Winkler}, \citenamefont {Strydom}, \citenamefont {Shimura}, \citenamefont
  {Sakakibara}, \citenamefont {Yu}, \citenamefont {Si},\ and\ \citenamefont
  {Paschen}}]{Custers2012}%
  \BibitemOpen
  \bibfield  {author} {\bibinfo {author} {\bibfnamefont {J.}~\bibnamefont
  {Custers}}, \bibinfo {author} {\bibfnamefont {K.}~\bibnamefont {Lorenzer}},
  \bibinfo {author} {\bibfnamefont {M.}~\bibnamefont {M{\"u}ller}}, \bibinfo
  {author} {\bibfnamefont {A.}~\bibnamefont {Prokofiev}}, \bibinfo {author}
  {\bibfnamefont {A.}~\bibnamefont {Sidorenko}}, \bibinfo {author}
  {\bibfnamefont {H.}~\bibnamefont {Winkler}}, \bibinfo {author} {\bibfnamefont
  {A.}~\bibnamefont {Strydom}}, \bibinfo {author} {\bibfnamefont
  {Y.}~\bibnamefont {Shimura}}, \bibinfo {author} {\bibfnamefont
  {T.}~\bibnamefont {Sakakibara}}, \bibinfo {author} {\bibfnamefont
  {R.}~\bibnamefont {Yu}}, \bibinfo {author} {\bibfnamefont {Q.}~\bibnamefont
  {Si}}, \ and\ \bibinfo {author} {\bibfnamefont {S.}~\bibnamefont {Paschen}},\
  }\href@noop {} {\bibfield  {journal} {\bibinfo  {journal} {Nature Materials}\
  }\textbf {\bibinfo {volume} {11}},\ \bibinfo {pages} {189} (\bibinfo {year}
  {2012})}\BibitemShut {NoStop}%
\bibitem [{\citenamefont {{D\"{o}nni}}\ \emph {et~al.}(1996)\citenamefont
  {{D\"{o}nni}}, \citenamefont {Ehlers}, \citenamefont {Maletta}, \citenamefont
  {Fischer}, \citenamefont {Kitazawa},\ and\ \citenamefont
  {Zolliker}}]{Donni1996}%
  \BibitemOpen
  \bibfield  {author} {\bibinfo {author} {\bibfnamefont {A.}~\bibnamefont
  {{D\"{o}nni}}}, \bibinfo {author} {\bibfnamefont {G.}~\bibnamefont {Ehlers}},
  \bibinfo {author} {\bibfnamefont {H.}~\bibnamefont {Maletta}}, \bibinfo
  {author} {\bibfnamefont {P.}~\bibnamefont {Fischer}}, \bibinfo {author}
  {\bibfnamefont {H.}~\bibnamefont {Kitazawa}}, \ and\ \bibinfo {author}
  {\bibfnamefont {M.}~\bibnamefont {Zolliker}},\ }\href@noop {} {\bibfield
  {journal} {\bibinfo  {journal} {J. Phys.: Cond. Mat.}\ }\textbf {\bibinfo
  {volume} {8}},\ \bibinfo {pages} {11213} (\bibinfo {year}
  {1996})}\BibitemShut {NoStop}%
\bibitem [{\citenamefont {Lee}(2008)}]{Lee2008}%
  \BibitemOpen
  \bibfield  {author} {\bibinfo {author} {\bibfnamefont {P.~A.}\ \bibnamefont
  {Lee}},\ }\href@noop {} {\bibfield  {journal} {\bibinfo  {journal} {Science}\
  }\textbf {\bibinfo {volume} {321}},\ \bibinfo {pages} {1306} (\bibinfo {year}
  {2008})}\BibitemShut {NoStop}%
\bibitem [{\citenamefont {Si}(2006)}]{Si2006}%
  \BibitemOpen
  \bibfield  {author} {\bibinfo {author} {\bibfnamefont {Q.}~\bibnamefont
  {Si}},\ }\href@noop {} {\bibfield  {journal} {\bibinfo  {journal} {Physica
  B}\ }\textbf {\bibinfo {volume} {378-380}},\ \bibinfo {pages} {23} (\bibinfo
  {year} {2006})}\BibitemShut {NoStop}%
\bibitem [{\citenamefont {Vojta}(2008)}]{Vojta2008}%
  \BibitemOpen
  \bibfield  {author} {\bibinfo {author} {\bibfnamefont {M.}~\bibnamefont
  {Vojta}},\ }\href@noop {} {\bibfield  {journal} {\bibinfo  {journal} {Phys.
  Rev. B}\ }\textbf {\bibinfo {volume} {78}},\ \bibinfo {pages} {125109}
  (\bibinfo {year} {2008})}\BibitemShut {NoStop}%
\bibitem [{\citenamefont {Coleman}\ and\ \citenamefont
  {Nevidomskyy}(2010)}]{Coleman2010}%
  \BibitemOpen
  \bibfield  {author} {\bibinfo {author} {\bibfnamefont {P.}~\bibnamefont
  {Coleman}}\ and\ \bibinfo {author} {\bibfnamefont {A.}~\bibnamefont
  {Nevidomskyy}},\ }\href@noop {} {\bibfield  {journal} {\bibinfo  {journal}
  {J. Low Temp. Phys.}\ }\textbf {\bibinfo {volume} {161}},\ \bibinfo {pages}
  {182} (\bibinfo {year} {2010})}\BibitemShut {NoStop}%
\bibitem [{\citenamefont {Kim}\ and\ \citenamefont {Aronson}(2011)}]{Kim2011}%
  \BibitemOpen
  \bibfield  {author} {\bibinfo {author} {\bibfnamefont {M.~S.}\ \bibnamefont
  {Kim}}\ and\ \bibinfo {author} {\bibfnamefont {M.~C.}\ \bibnamefont
  {Aronson}},\ }\href@noop {} {\bibfield  {journal} {\bibinfo  {journal} {J.
  Phys.: Cond. Mat.}\ }\textbf {\bibinfo {volume} {23}},\ \bibinfo {pages}
  {164204} (\bibinfo {year} {2011})}\BibitemShut {NoStop}%
\bibitem [{\citenamefont {Bud\^{O}\c{C}\"{O}ko}\ \emph {et~al.}(2004)\citenamefont
  {Bud\^{O}\c{C}\"{O}ko}, \citenamefont {Morosan},\ and\ \citenamefont
  {Canfield}}]{Budko2004}%
  \BibitemOpen
  \bibfield  {author} {\bibinfo {author} {\bibfnamefont {S.~L.}\ \bibnamefont
  {Bud\^{O}\c{C}\"{O}ko}}, \bibinfo {author} {\bibfnamefont {E.}~\bibnamefont {Morosan}}, \
  and\ \bibinfo {author} {\bibfnamefont {P.~C.}\ \bibnamefont {Canfield}},\
  }\href@noop {} {\bibfield  {journal} {\bibinfo  {journal} {Phys. Rev. B}\
  }\textbf {\bibinfo {volume} {69}},\ \bibinfo {pages} {014415} (\bibinfo
  {year} {2004})}\BibitemShut {NoStop}%
\bibitem [{\citenamefont {Kim}\ and\ \citenamefont {Aronson}(2013)}]{Kim2013}%
  \BibitemOpen
  \bibfield  {author} {\bibinfo {author} {\bibfnamefont {M.~S.}\ \bibnamefont
  {Kim}}\ and\ \bibinfo {author} {\bibfnamefont {M.~C.}\ \bibnamefont
  {Aronson}},\ }\href@noop {} {\bibfield  {journal} {\bibinfo  {journal} {Phys.
  Rev. Lett.}\ }\textbf {\bibinfo {volume} {110}},\ \bibinfo {pages} {017201}
  (\bibinfo {year} {2013})}\BibitemShut {NoStop}%
\bibitem [{\citenamefont {Oyamada}\ \emph {et~al.}(1996)\citenamefont
  {Oyamada}, \citenamefont {Kamioka}, \citenamefont {Hashi}, \citenamefont
  {Maegawa}, \citenamefont {Goto},\ and\ \citenamefont
  {Kitzawa}}]{Oyamada1996}%
  \BibitemOpen
  \bibfield  {author} {\bibinfo {author} {\bibfnamefont {A.}~\bibnamefont
  {Oyamada}}, \bibinfo {author} {\bibfnamefont {K.}~\bibnamefont {Kamioka}},
  \bibinfo {author} {\bibfnamefont {K.}~\bibnamefont {Hashi}}, \bibinfo
  {author} {\bibfnamefont {S.}~\bibnamefont {Maegawa}}, \bibinfo {author}
  {\bibfnamefont {T.}~\bibnamefont {Goto}}, \ and\ \bibinfo {author}
  {\bibfnamefont {H.}~\bibnamefont {Kitzawa}},\ }\href@noop {} {\bibfield
  {journal} {\bibinfo  {journal} {J. Phys. Soc. Jpn.}\ }\textbf {\bibinfo
  {volume} {65}},\ \bibinfo {pages} {128} (\bibinfo {year} {1996})}\BibitemShut
  {NoStop}%
\bibitem [{\citenamefont {Isikawa}\ \emph {et~al.}(1996)\citenamefont
  {Isikawa}, \citenamefont {Mizushima}, \citenamefont {Fukushima},
  \citenamefont {Kuwai}, \citenamefont {Sakurai},\ and\ \citenamefont
  {Kitzawa}}]{Isikawa1996}%
  \BibitemOpen
  \bibfield  {author} {\bibinfo {author} {\bibfnamefont {Y.}~\bibnamefont
  {Isikawa}}, \bibinfo {author} {\bibfnamefont {T.}~\bibnamefont {Mizushima}},
  \bibinfo {author} {\bibfnamefont {N.}~\bibnamefont {Fukushima}}, \bibinfo
  {author} {\bibfnamefont {T.}~\bibnamefont {Kuwai}}, \bibinfo {author}
  {\bibfnamefont {J.}~\bibnamefont {Sakurai}}, \ and\ \bibinfo {author}
  {\bibfnamefont {H.}~\bibnamefont {Kitzawa}},\ }\href@noop {} {\bibfield
  {journal} {\bibinfo  {journal} {J. Phys. Soc. Jpn.}\ }\textbf {\bibinfo
  {volume} {65 {\normalfont Suppl. B}}},\ \bibinfo {pages} {117} (\bibinfo
  {year} {1996})}\BibitemShut {NoStop}%
\bibitem [{\citenamefont {Goto}\ \emph {et~al.}(2002)\citenamefont {Goto},
  \citenamefont {Hane}, \citenamefont {Umeo}, \citenamefont {Takabatake},\ and\
  \citenamefont {Isikawa}}]{Goto2002}%
  \BibitemOpen
  \bibfield  {author} {\bibinfo {author} {\bibfnamefont {T.}~\bibnamefont
  {Goto}}, \bibinfo {author} {\bibfnamefont {S.}~\bibnamefont {Hane}}, \bibinfo
  {author} {\bibfnamefont {K.}~\bibnamefont {Umeo}}, \bibinfo {author}
  {\bibfnamefont {T.}~\bibnamefont {Takabatake}}, \ and\ \bibinfo {author}
  {\bibfnamefont {Y.}~\bibnamefont {Isikawa}},\ }\href@noop {} {\bibfield
  {journal} {\bibinfo  {journal} {J. Phys. Chem. Sol.}\ }\textbf {\bibinfo
  {volume} {63}},\ \bibinfo {pages} {1159} (\bibinfo {year}
  {2002})}\BibitemShut {NoStop}%
\bibitem [{\citenamefont {Proke{\v{s}}}\ \emph {et~al.}(2006)\citenamefont
  {Proke{\v{s}}}, \citenamefont {Manuel}, \citenamefont {Adroja}, \citenamefont
  {Kitazawa}, \citenamefont {Goto},\ and\ \citenamefont
  {Isikawa}}]{Prokevs2006}%
  \BibitemOpen
  \bibfield  {author} {\bibinfo {author} {\bibfnamefont {K.}~\bibnamefont
  {Proke{\v{s}}}}, \bibinfo {author} {\bibfnamefont {P.}~\bibnamefont
  {Manuel}}, \bibinfo {author} {\bibfnamefont {D.}~\bibnamefont {Adroja}},
  \bibinfo {author} {\bibfnamefont {H.}~\bibnamefont {Kitazawa}}, \bibinfo
  {author} {\bibfnamefont {T.}~\bibnamefont {Goto}}, \ and\ \bibinfo {author}
  {\bibfnamefont {Y.}~\bibnamefont {Isikawa}},\ }\href@noop {} {\bibfield
  {journal} {\bibinfo  {journal} {Physica B}\ }\textbf {\bibinfo {volume}
  {385}},\ \bibinfo {pages} {359} (\bibinfo {year} {2006})}\BibitemShut
  {NoStop}%
\bibitem [{\citenamefont {Oyamada}\ \emph {et~al.}(2008)\citenamefont
  {Oyamada}, \citenamefont {Maegawa}, \citenamefont {Nishiyama}, \citenamefont
  {Kitazawa},\ and\ \citenamefont {Isikawa}}]{Oyamada2008}%
  \BibitemOpen
  \bibfield  {author} {\bibinfo {author} {\bibfnamefont {A.}~\bibnamefont
  {Oyamada}}, \bibinfo {author} {\bibfnamefont {S.}~\bibnamefont {Maegawa}},
  \bibinfo {author} {\bibfnamefont {M.}~\bibnamefont {Nishiyama}}, \bibinfo
  {author} {\bibfnamefont {H.}~\bibnamefont {Kitazawa}}, \ and\ \bibinfo
  {author} {\bibfnamefont {Y.}~\bibnamefont {Isikawa}},\ }\href@noop {}
  {\bibfield  {journal} {\bibinfo  {journal} {Phys. Rev. B}\ }\textbf {\bibinfo
  {volume} {77}},\ \bibinfo {pages} {064432} (\bibinfo {year}
  {2008})}\BibitemShut {NoStop}%
\bibitem [{\citenamefont {Villars}\ and\ \citenamefont
  {Cenzual}()}]{Pearsons2012}%
  \BibitemOpen
  \bibfield  {author} {\bibinfo {author} {\bibfnamefont {P.}~\bibnamefont
  {Villars}}\ and\ \bibinfo {author} {\bibfnamefont {K.}~\bibnamefont
  {Cenzual}},\ }\href@noop {} {\enquote {\bibinfo {title} {{Pearson's Crystal
  Data - Crystal Structure Database for Inorganic Compounds (on CD-Rom),
  {\normalfont Release 2011/12}}},}\ }\bibinfo {note} {{ASM International,
  Materials Park, Ohio, USA}}\BibitemShut {NoStop}%
\bibitem [{\citenamefont {N{\'u}{\~n}ez-Regueiro}\ \emph
  {et~al.}(1997)\citenamefont {N{\'u}{\~n}ez-Regueiro}, \citenamefont
  {Lacroix},\ and\ \citenamefont {Canals}}]{Nunez-Regueiro1997}%
  \BibitemOpen
  \bibfield  {author} {\bibinfo {author} {\bibfnamefont {M.}~\bibnamefont
  {N{\'u}{\~n}ez-Regueiro}}, \bibinfo {author} {\bibfnamefont {C.}~\bibnamefont
  {Lacroix}}, \ and\ \bibinfo {author} {\bibfnamefont {B.}~\bibnamefont
  {Canals}},\ }\href@noop {} {\bibfield  {journal} {\bibinfo  {journal}
  {Physica C}\ }\textbf {\bibinfo {volume} {282}},\ \bibinfo {pages} {1885}
  (\bibinfo {year} {1997})}\BibitemShut {NoStop}%
\bibitem [{\citenamefont {Motome}\ \emph {et~al.}(2010)\citenamefont {Motome},
  \citenamefont {Nakamikawa}, \citenamefont {Yamaji},\ and\ \citenamefont
  {Udagawa}}]{Motome2010}%
  \BibitemOpen
  \bibfield  {author} {\bibinfo {author} {\bibfnamefont {Y.}~\bibnamefont
  {Motome}}, \bibinfo {author} {\bibfnamefont {K.}~\bibnamefont {Nakamikawa}},
  \bibinfo {author} {\bibfnamefont {Y.}~\bibnamefont {Yamaji}}, \ and\ \bibinfo
  {author} {\bibfnamefont {M.}~\bibnamefont {Udagawa}},\ }\href@noop {}
  {\bibfield  {journal} {\bibinfo  {journal} {Phys. Rev. Lett.}\ }\textbf
  {\bibinfo {volume} {105}},\ \bibinfo {pages} {036403} (\bibinfo {year}
  {2010})}\BibitemShut {NoStop}%
\bibitem [{\citenamefont {Isikawa}\ \emph {et~al.}(2000)\citenamefont
  {Isikawa}, \citenamefont {Kuwai}, \citenamefont {Mizushima}, \citenamefont
  {Abe}, \citenamefont {Nakamura},\ and\ \citenamefont
  {Sakurai}}]{Isikawa2000}%
  \BibitemOpen
  \bibfield  {author} {\bibinfo {author} {\bibfnamefont {Y.}~\bibnamefont
  {Isikawa}}, \bibinfo {author} {\bibfnamefont {T.}~\bibnamefont {Kuwai}},
  \bibinfo {author} {\bibfnamefont {T.}~\bibnamefont {Mizushima}}, \bibinfo
  {author} {\bibfnamefont {T.}~\bibnamefont {Abe}}, \bibinfo {author}
  {\bibfnamefont {G.}~\bibnamefont {Nakamura}}, \ and\ \bibinfo {author}
  {\bibfnamefont {J.}~\bibnamefont {Sakurai}},\ }\href@noop {} {\bibfield
  {journal} {\bibinfo  {journal} {Physica B}\ }\textbf {\bibinfo {volume}
  {281}},\ \bibinfo {pages} {365} (\bibinfo {year} {2000})}\BibitemShut
  {NoStop}%
\bibitem [{\citenamefont {Fritsch}\ \emph {et~al.}(2013)\citenamefont
  {Fritsch}, \citenamefont {Huang}, \citenamefont {Bagrets}, \citenamefont
  {Grube}, \citenamefont {Schumann},\ and\ \citenamefont {{v.
  L{\"o}hneysen}}}]{Fritsch2013}%
  \BibitemOpen
  \bibfield  {author} {\bibinfo {author} {\bibfnamefont {V.}~\bibnamefont
  {Fritsch}}, \bibinfo {author} {\bibfnamefont {C.-L.}\ \bibnamefont {Huang}},
  \bibinfo {author} {\bibfnamefont {N.}~\bibnamefont {Bagrets}}, \bibinfo
  {author} {\bibfnamefont {K.}~\bibnamefont {Grube}}, \bibinfo {author}
  {\bibfnamefont {S.}~\bibnamefont {Schumann}}, \ and\ \bibinfo {author}
  {\bibfnamefont {H.}~\bibnamefont {{v. L{\"o}hneysen}}},\ }\href@noop {}
  {\bibfield  {journal} {\bibinfo  {journal} {phys. stat. sol. (b)}\ }\textbf
  {\bibinfo {volume} {250}},\ \bibinfo {pages} {506 } (\bibinfo {year}
  {2013})}\BibitemShut {NoStop}%
\bibitem [{ame()}]{ames}%
  \BibitemOpen
  \href@noop {} {}\bibinfo {note} {{High-purity rare-earth metals acquired from
  Materials Preparation Center, Ames Laboratory, US DOE Basic Energy Sciences,
  Ames, IA, USA, $<$http://www.mpc.ameslab.gov$>$}}\BibitemShut {NoStop}%
\bibitem [{\citenamefont {Gribanov}\ \emph {et~al.}(2006)\citenamefont
  {Gribanov}, \citenamefont {Tursina}, \citenamefont {Murashova}, \citenamefont
  {Seropegin}, \citenamefont {Bauer}, \citenamefont {Kaldarar}, \citenamefont
  {Lackner}, \citenamefont {Michor}, \citenamefont {Royanian}, \citenamefont
  {Reissner},\ and\ \citenamefont {Rogl}}]{Gribanov2006}%
  \BibitemOpen
  \bibfield  {author} {\bibinfo {author} {\bibfnamefont {A.}~\bibnamefont
  {Gribanov}}, \bibinfo {author} {\bibfnamefont {A.}~\bibnamefont {Tursina}},
  \bibinfo {author} {\bibfnamefont {E.}~\bibnamefont {Murashova}}, \bibinfo
  {author} {\bibfnamefont {Y.}~\bibnamefont {Seropegin}}, \bibinfo {author}
  {\bibfnamefont {E.}~\bibnamefont {Bauer}}, \bibinfo {author} {\bibfnamefont
  {H.}~\bibnamefont {Kaldarar}}, \bibinfo {author} {\bibfnamefont
  {R.}~\bibnamefont {Lackner}}, \bibinfo {author} {\bibfnamefont
  {H.}~\bibnamefont {Michor}}, \bibinfo {author} {\bibfnamefont
  {E.}~\bibnamefont {Royanian}}, \bibinfo {author} {\bibfnamefont
  {M.}~\bibnamefont {Reissner}}, \ and\ \bibinfo {author} {\bibfnamefont
  {P.}~\bibnamefont {Rogl}},\ }\href@noop {} {\bibfield  {journal} {\bibinfo
  {journal} {J. Phys.: Cond. Mat.}\ }\textbf {\bibinfo {volume} {18}},\
  \bibinfo {pages} {9593} (\bibinfo {year} {2006})}\BibitemShut {NoStop}%
\bibitem [{\citenamefont {Schank}\ \emph {et~al.}(1994)\citenamefont {Schank},
  \citenamefont {J{\"a}hrling}, \citenamefont {Luo}, \citenamefont {Grauel},
  \citenamefont {Wassilew}, \citenamefont {Borth}, \citenamefont {Olesch},
  \citenamefont {Bredl}, \citenamefont {Geibel},\ and\ \citenamefont
  {Steglich}}]{Schank1994}%
  \BibitemOpen
  \bibfield  {author} {\bibinfo {author} {\bibfnamefont {C.}~\bibnamefont
  {Schank}}, \bibinfo {author} {\bibfnamefont {F.}~\bibnamefont
  {J{\"a}hrling}}, \bibinfo {author} {\bibfnamefont {L.}~\bibnamefont {Luo}},
  \bibinfo {author} {\bibfnamefont {A.}~\bibnamefont {Grauel}}, \bibinfo
  {author} {\bibfnamefont {C.}~\bibnamefont {Wassilew}}, \bibinfo {author}
  {\bibfnamefont {R.}~\bibnamefont {Borth}}, \bibinfo {author} {\bibfnamefont
  {G.}~\bibnamefont {Olesch}}, \bibinfo {author} {\bibfnamefont
  {C.}~\bibnamefont {Bredl}}, \bibinfo {author} {\bibfnamefont
  {C.}~\bibnamefont {Geibel}}, \ and\ \bibinfo {author} {\bibfnamefont
  {F.}~\bibnamefont {Steglich}},\ }\href@noop {} {\bibfield  {journal}
  {\bibinfo  {journal} {J. All. \& Comp.}\ }\textbf {\bibinfo {volume} {207}},\
  \bibinfo {pages} {329} (\bibinfo {year} {1994})}\BibitemShut {NoStop}%
\bibitem [{\citenamefont {v.~L{\"o}hneysen}(1996)}]{Lohneysen1996}%
  \BibitemOpen
  \bibfield  {author} {\bibinfo {author} {\bibfnamefont {H.}~\bibnamefont
  {v.~L{\"o}hneysen}},\ }\href@noop {} {\bibfield  {journal} {\bibinfo
  {journal} {J.~Phys.: Condens. Matter}\ }\textbf {\bibinfo {volume} {8}},\
  \bibinfo {pages} {9689 } (\bibinfo {year} {1996})}\BibitemShut {NoStop}%
\bibitem [{\citenamefont {Grimvall}(1999)}]{Grimvall1999}%
  \BibitemOpen
  \bibfield  {author} {\bibinfo {author} {\bibfnamefont {G.}~\bibnamefont
  {Grimvall}},\ }\href@noop {} {\emph {\bibinfo {title} {Thermophysical
  properties of materials}}}\ (\bibinfo  {publisher} {The Royal Institute of
  Technology},\ \bibinfo {address} {Stockholm, Sweden},\ \bibinfo {year}
  {1999})\ \bibinfo {note} {{p.} 324 ff.}\BibitemShut {Stop}%
\bibitem [{\citenamefont {Tang}\ \emph {et~al.}(1996)\citenamefont {Tang},
  \citenamefont {Matsushita}, \citenamefont {Kitazawa},\ and\ \citenamefont
  {Matsumoto}}]{Tang1996}%
  \BibitemOpen
  \bibfield  {author} {\bibinfo {author} {\bibfnamefont {J.}~\bibnamefont
  {Tang}}, \bibinfo {author} {\bibfnamefont {A.}~\bibnamefont {Matsushita}},
  \bibinfo {author} {\bibfnamefont {H.}~\bibnamefont {Kitazawa}}, \ and\
  \bibinfo {author} {\bibfnamefont {T.}~\bibnamefont {Matsumoto}},\ }\href@noop
  {} {\bibfield  {journal} {\bibinfo  {journal} {Physica B}\ }\textbf {\bibinfo
  {volume} {217}},\ \bibinfo {pages} {97} (\bibinfo {year} {1996})}\BibitemShut
  {NoStop}%
\bibitem [{\citenamefont {Fritsch}\ \emph {et~al.}(2006)\citenamefont
  {Fritsch}, \citenamefont {Thompson}, \citenamefont {Sarrao}, \citenamefont
  {{Krug von Nidda}}, \citenamefont {Eremina},\ and\ \citenamefont
  {Loidl}}]{Fritsch2006b}%
  \BibitemOpen
  \bibfield  {author} {\bibinfo {author} {\bibfnamefont {V.}~\bibnamefont
  {Fritsch}}, \bibinfo {author} {\bibfnamefont {J.~D.}\ \bibnamefont
  {Thompson}}, \bibinfo {author} {\bibfnamefont {J.~L.}\ \bibnamefont
  {Sarrao}}, \bibinfo {author} {\bibfnamefont {H.-A.}\ \bibnamefont {{Krug von
  Nidda}}}, \bibinfo {author} {\bibfnamefont {R.~M.}\ \bibnamefont {Eremina}},
  \ and\ \bibinfo {author} {\bibfnamefont {A.}~\bibnamefont {Loidl}},\
  }\href@noop {} {\bibfield  {journal} {\bibinfo  {journal} {Phys. Rev. B}\
  }\textbf {\bibinfo {volume} {73}},\ \bibinfo {pages} {094413} (\bibinfo
  {year} {2006})}\BibitemShut {NoStop}%
\bibitem [{\citenamefont {Lakner}\ \emph {et~al.}(1994)\citenamefont {Lakner},
  \citenamefont {{v. L{\"o}hneysen}}, \citenamefont {Langenfeld},\ and\
  \citenamefont {W{\"o}lfle}}]{Lakner1994}%
  \BibitemOpen
  \bibfield  {author} {\bibinfo {author} {\bibfnamefont {M.}~\bibnamefont
  {Lakner}}, \bibinfo {author} {\bibfnamefont {H.}~\bibnamefont {{v.
  L{\"o}hneysen}}}, \bibinfo {author} {\bibfnamefont {A.}~\bibnamefont
  {Langenfeld}}, \ and\ \bibinfo {author} {\bibfnamefont {P.}~\bibnamefont
  {W{\"o}lfle}},\ }\href {\doibase 10.1103/PhysRevB.50.17064} {\bibfield
  {journal} {\bibinfo  {journal} {Phys. Rev. B}\ }\textbf {\bibinfo {volume}
  {50}},\ \bibinfo {pages} {17064} (\bibinfo {year} {1994})}\BibitemShut
  {NoStop}%
\bibitem [{\citenamefont {Miranda}\ \emph {et~al.}(1997)\citenamefont
  {Miranda}, \citenamefont {Dobrosavljevic},\ and\ \citenamefont
  {Kotliar}}]{Miranda1997}%
  \BibitemOpen
  \bibfield  {author} {\bibinfo {author} {\bibfnamefont {E.}~\bibnamefont
  {Miranda}}, \bibinfo {author} {\bibfnamefont {V.}~\bibnamefont
  {Dobrosavljevic}}, \ and\ \bibinfo {author} {\bibfnamefont {G.}~\bibnamefont
  {Kotliar}},\ }\href {\doibase 10.1103/PhysRevLett.78.290} {\bibfield
  {journal} {\bibinfo  {journal} {Phys. Rev. Lett.}\ }\textbf {\bibinfo
  {volume} {78}},\ \bibinfo {pages} {290} (\bibinfo {year} {1997})}\BibitemShut
  {NoStop}%
\bibitem [{\citenamefont {Bernal}\ \emph {et~al.}(1996)\citenamefont {Bernal},
  \citenamefont {MacLaughlin}, \citenamefont {Amato}, \citenamefont
  {Feyerherm}, \citenamefont {Gygax}, \citenamefont {Schenck}, \citenamefont
  {Heffner}, \citenamefont {Le}, \citenamefont {Nieuwenhuys}, \citenamefont
  {Andraka}, \citenamefont {L\"ohneysen}, \citenamefont {Stockert},\ and\
  \citenamefont {Ott}}]{Bernal1996}%
  \BibitemOpen
  \bibfield  {author} {\bibinfo {author} {\bibfnamefont {O.~O.}\ \bibnamefont
  {Bernal}}, \bibinfo {author} {\bibfnamefont {D.~E.}\ \bibnamefont
  {MacLaughlin}}, \bibinfo {author} {\bibfnamefont {A.}~\bibnamefont {Amato}},
  \bibinfo {author} {\bibfnamefont {R.}~\bibnamefont {Feyerherm}}, \bibinfo
  {author} {\bibfnamefont {F.~N.}\ \bibnamefont {Gygax}}, \bibinfo {author}
  {\bibfnamefont {A.}~\bibnamefont {Schenck}}, \bibinfo {author} {\bibfnamefont
  {R.~H.}\ \bibnamefont {Heffner}}, \bibinfo {author} {\bibfnamefont {L.~P.}\
  \bibnamefont {Le}}, \bibinfo {author} {\bibfnamefont {G.~J.}\ \bibnamefont
  {Nieuwenhuys}}, \bibinfo {author} {\bibfnamefont {B.}~\bibnamefont
  {Andraka}}, \bibinfo {author} {\bibfnamefont {H.~v.}\ \bibnamefont
  {L\"ohneysen}}, \bibinfo {author} {\bibfnamefont {O.}~\bibnamefont
  {Stockert}}, \ and\ \bibinfo {author} {\bibfnamefont {H.~R.}\ \bibnamefont
  {Ott}},\ }\href {\doibase 10.1103/PhysRevB.54.13000} {\bibfield  {journal}
  {\bibinfo  {journal} {Phys. Rev. B}\ }\textbf {\bibinfo {volume} {54}},\
  \bibinfo {pages} {13000} (\bibinfo {year} {1996})}\BibitemShut {NoStop}%
\bibitem [{\citenamefont {Hamann}\ \emph {et~al.}(2013)\citenamefont {Hamann},
  \citenamefont {Stockert}, \citenamefont {Fritsch}, \citenamefont {Grube},
  \citenamefont {Schneidewind},\ and\ \citenamefont {{v.
  L\"ohneysen}}}]{Hamann2013}%
  \BibitemOpen
  \bibfield  {author} {\bibinfo {author} {\bibfnamefont {A.}~\bibnamefont
  {Hamann}}, \bibinfo {author} {\bibfnamefont {O.}~\bibnamefont {Stockert}},
  \bibinfo {author} {\bibfnamefont {V.}~\bibnamefont {Fritsch}}, \bibinfo
  {author} {\bibfnamefont {K.}~\bibnamefont {Grube}}, \bibinfo {author}
  {\bibfnamefont {A.}~\bibnamefont {Schneidewind}}, \ and\ \bibinfo {author}
  {\bibfnamefont {H.}~\bibnamefont {{v. L\"ohneysen}}},\ }\href@noop {}
  {\bibfield  {journal} {\bibinfo  {journal} {Phys. Rev. Lett.}\ }\textbf
  {\bibinfo {volume} {110}},\ \bibinfo {pages} {096404} (\bibinfo {year}
  {2013})}\BibitemShut {NoStop}%
\bibitem [{\citenamefont {Schmalian}\ and\ \citenamefont
  {Batista}(2008)}]{Schmalian2008}%
  \BibitemOpen
  \bibfield  {author} {\bibinfo {author} {\bibfnamefont {J.}~\bibnamefont
  {Schmalian}}\ and\ \bibinfo {author} {\bibfnamefont {C.~D.}\ \bibnamefont
  {Batista}},\ }\href@noop {} {\bibfield  {journal} {\bibinfo  {journal} {Phys.
  Rev. B}\ }\textbf {\bibinfo {volume} {77}},\ \bibinfo {pages} {094406}
  (\bibinfo {year} {2008})}\BibitemShut {NoStop}%
\bibitem [{\citenamefont {Sebastian}\ \emph {et~al.}(2006)\citenamefont
  {Sebastian}, \citenamefont {Harrison}, \citenamefont {Batista}, \citenamefont
  {Balicas}, \citenamefont {Jaime}, \citenamefont {Sharma}, \citenamefont
  {Kawashima},\ and\ \citenamefont {Fisher}}]{Sebastian2006}%
  \BibitemOpen
  \bibfield  {author} {\bibinfo {author} {\bibfnamefont {S.~E.}\ \bibnamefont
  {Sebastian}}, \bibinfo {author} {\bibfnamefont {N.}~\bibnamefont {Harrison}},
  \bibinfo {author} {\bibfnamefont {C.~D.}\ \bibnamefont {Batista}}, \bibinfo
  {author} {\bibfnamefont {L.}~\bibnamefont {Balicas}}, \bibinfo {author}
  {\bibfnamefont {M.}~\bibnamefont {Jaime}}, \bibinfo {author} {\bibfnamefont
  {P.~A.}\ \bibnamefont {Sharma}}, \bibinfo {author} {\bibfnamefont
  {N.}~\bibnamefont {Kawashima}}, \ and\ \bibinfo {author} {\bibfnamefont
  {I.~R.}\ \bibnamefont {Fisher}},\ }\href@noop {} {\bibfield  {journal}
  {\bibinfo  {journal} {Nature}\ }\textbf {\bibinfo {volume} {441}},\ \bibinfo
  {pages} {617} (\bibinfo {year} {2006})}\BibitemShut {NoStop}%
\bibitem [{\citenamefont {Batista}\ \emph {et~al.}(2007)\citenamefont
  {Batista}, \citenamefont {Schmalian}, \citenamefont {Kawashima},
  \citenamefont {Sengupta}, \citenamefont {Sebastian}, \citenamefont
  {Harrison}, \citenamefont {Jaime},\ and\ \citenamefont
  {Fisher}}]{Batista2007}%
  \BibitemOpen
  \bibfield  {author} {\bibinfo {author} {\bibfnamefont {C.~D.}\ \bibnamefont
  {Batista}}, \bibinfo {author} {\bibfnamefont {J.}~\bibnamefont {Schmalian}},
  \bibinfo {author} {\bibfnamefont {N.}~\bibnamefont {Kawashima}}, \bibinfo
  {author} {\bibfnamefont {P.}~\bibnamefont {Sengupta}}, \bibinfo {author}
  {\bibfnamefont {S.}~\bibnamefont {Sebastian}}, \bibinfo {author}
  {\bibfnamefont {N.}~\bibnamefont {Harrison}}, \bibinfo {author}
  {\bibfnamefont {M.}~\bibnamefont {Jaime}}, \ and\ \bibinfo {author}
  {\bibfnamefont {I.}~\bibnamefont {Fisher}},\ }\href@noop {} {\bibfield
  {journal} {\bibinfo  {journal} {Phys. Rev. Lett.}\ }\textbf {\bibinfo
  {volume} {98}},\ \bibinfo {pages} {257201} (\bibinfo {year}
  {2007})}\BibitemShut {NoStop}%
\bibitem [{\citenamefont {Abrahams}\ and\ \citenamefont
  {W{\"o}lfle}(2012)}]{Abrahams2012}%
  \BibitemOpen
  \bibfield  {author} {\bibinfo {author} {\bibfnamefont {E.}~\bibnamefont
  {Abrahams}}\ and\ \bibinfo {author} {\bibfnamefont {P.}~\bibnamefont
  {W{\"o}lfle}},\ }\href@noop {} {\bibfield  {journal} {\bibinfo  {journal}
  {Proc. Nat. Acad. Sci.}\ }\textbf {\bibinfo {volume} {109}},\ \bibinfo
  {pages} {3238} (\bibinfo {year} {2012})}\BibitemShut {NoStop}%
\bibitem [{\citenamefont {Helton}\ \emph {et~al.}(2007)\citenamefont {Helton},
  \citenamefont {Matan}, \citenamefont {Shores}, \citenamefont {Nytko},
  \citenamefont {Bartlett}, \citenamefont {Yoshida}, \citenamefont {Takano},
  \citenamefont {Suslov}, \citenamefont {Qiu}, \citenamefont {Chung},
  \citenamefont {Nocera},\ and\ \citenamefont {Lee}}]{Helton2007}%
  \BibitemOpen
  \bibfield  {author} {\bibinfo {author} {\bibfnamefont {J.~S.}\ \bibnamefont
  {Helton}}, \bibinfo {author} {\bibfnamefont {K.}~\bibnamefont {Matan}},
  \bibinfo {author} {\bibfnamefont {M.~P.}\ \bibnamefont {Shores}}, \bibinfo
  {author} {\bibfnamefont {E.~A.}\ \bibnamefont {Nytko}}, \bibinfo {author}
  {\bibfnamefont {B.~M.}\ \bibnamefont {Bartlett}}, \bibinfo {author}
  {\bibfnamefont {Y.}~\bibnamefont {Yoshida}}, \bibinfo {author} {\bibfnamefont
  {Y.}~\bibnamefont {Takano}}, \bibinfo {author} {\bibfnamefont
  {A.}~\bibnamefont {Suslov}}, \bibinfo {author} {\bibfnamefont
  {Y.}~\bibnamefont {Qiu}}, \bibinfo {author} {\bibfnamefont {J.-H.}\
  \bibnamefont {Chung}}, \bibinfo {author} {\bibfnamefont {D.~G.}\ \bibnamefont
  {Nocera}}, \ and\ \bibinfo {author} {\bibfnamefont {Y.~S.}\ \bibnamefont
  {Lee}},\ }\href {\doibase 10.1103/PhysRevLett.98.107204} {\bibfield
  {journal} {\bibinfo  {journal} {Phys. Rev. Lett.}\ }\textbf {\bibinfo
  {volume} {98}},\ \bibinfo {pages} {107204} (\bibinfo {year}
  {2007})}\BibitemShut {NoStop}%
\bibitem [{\citenamefont {Senthil}\ \emph {et~al.}(2004)\citenamefont
  {Senthil}, \citenamefont {Vojta},\ and\ \citenamefont
  {Sachdev}}]{Senthil2004}%
  \BibitemOpen
  \bibfield  {author} {\bibinfo {author} {\bibfnamefont {T.}~\bibnamefont
  {Senthil}}, \bibinfo {author} {\bibfnamefont {M.}~\bibnamefont {Vojta}}, \
  and\ \bibinfo {author} {\bibfnamefont {S.}~\bibnamefont {Sachdev}},\ }\href
  {\doibase 10.1103/PhysRevB.69.035111} {\bibfield  {journal} {\bibinfo
  {journal} {Phys. Rev. B}\ }\textbf {\bibinfo {volume} {69}},\ \bibinfo
  {pages} {035111} (\bibinfo {year} {2004})}\BibitemShut {NoStop}%
\end{thebibliography}
\end{document}